\documentclass{aa}
\usepackage{graphicx}
\usepackage{natbib}
\usepackage{txfonts}
\begin{document}
   \title{A Radiation Driven Implosion Model\\
   for the Enhanced Luminosity of Protostars near HII Regions}

   \subtitle{}

   \author{K. Motoyama\inst{1}
           \and
           T. Umemoto\inst{2}
       \and
       H. Shang\inst{3}
      }


   \institute{Theoretical Institute for Advanced Research in
     Astrophysics, Dept. of Physics, National Tsing Hua University, 101,
     Sec. 2, Kuang-Fu Rd., Hsin-Chu, 30013, Taiwan.
     \and
     Nobeyama Radio Observatory, Nobeyama, Minamimaki,
              Minamisaku, Nagano 384-1305, Japan.
          \and
          Institute of Astronomy and Astrophysics, Academia Sinica,
     Taipei 106, Taiwan.
             }

   \date{Received December 29, 2006; accepted February 7, 2007}


\abstract
   {Molecular clouds near the \ion{H}{ii} regions tend to harbor more luminous
   protostars.}
   {Our aim in this paper is to investigate whether or not
   radiation-driven implosion mechanism enhances luminosity of protostars
   near regions of high-ionizing fluxes.
   }
   {We performed numerical simulations to model collapse of cores
   exposed to UV radiation from O stars. We investigated dependence
   of mass loss rates on the initial density profiles of cores and
   variation of UV fluxes. We derived simple analytic
   estimates of
   accretion rates and final masses of protostars.}
   {Radiation-driven implosion mechanism can increase accretion
   rates of protostars by 1-2 orders of magnitude. On the other
   hand, mass loss due to photo-evaporation is not large enough to
   have
   a significant impact  on the luminosity. The increase of accretion
   rate makes luminosity 1-2 orders higher than those of protostars
   that form without external triggering.}
   {Radiation-driven implosion can help explain the observed
   higher luminosity of protostars in molecular clouds near
   \ion{H}{ii} regions.}


   \maketitle
%

\section{Introduction}
Massive stars play a very important role in star formation history
in giant molecular clouds. Strong radiation from massive stars
impacts on surrounding environment, and may trigger star formation
nearby. Two scenarios of triggered star formation near \ion{H}{ii}
regions have been suggested. One of them is  ``collect and
collapse'' \citep{1977ApJ...214..725E, 1994MNRAS.268..291W}. As the
\ion{H}{ii} region expands, it pushes ambient gas into a shell.
Subsequent fragmentation and collapse of the swept-up shell may form
the next generation of stars. For example, a core within one
fragment of the molecular ring surrounding Galactic \ion{H}{ii}
region Sh104 contains a young cluster \citep{2003A&A...408L..25D}.
Radiation-driven implosion (hereafter, RDI) is the other scenario
\citep{1989ApJ...346..735B, 1990ApJ...354..529B}. When an expanding
\ion{H}{ii} region engulfs a cloud core, the strong UV radiation
will compress the core. Bright-rimmed clouds of cometary shapes
found at the edges of relatively old \ion{H}{ii} regions, are
explained by \citet{1994A&A...289..559L} with a RDI model. To
understand star formation in giant molecular clouds, we study
effects of the RDI.

Observational results indicate that the \ion{H}{ii} regions can
strongly affect star formation nearby. \citet{2001PASJ...53...85D}
investigated the maximum luminosity of protostars as a function of
the parent cloud mass. They found that molecular clouds near
\ion{H}{ii} regions contain more luminous protostars than others.
Moreover, \citet{1989ApJ...342L..87S} showed that the ratios of
luminosity of protostar to core mass of bright-rimmed globules are
much higher than those in dark globules. These results indicate that
the RDI affects the luminosity of protostars near \ion{H}{ii}
region.

Recent studies suggest that triggered star formation by external
compression can increase accretion rates and luminosity of
protostars. \citet{2003MNRAS.344..461M} investigated collapses of
cores compressed by external shock waves and found that accretion
rates were enhanced. Some observations support this result.
\citet{2006A&A...453..145B} observed the Class 0 protostar IRAS 4A
in star formation cloud NGC1333, whose star formation is suspected
to be triggered by powerful molecular outflows
\citep{2001ApJ...546L..49S}. They showed a numerical model of
collapse triggered by external pressure, which reproduces the
observed density and velocity profiles. Moreover, they deduced a
high accretion rate of $(0.7-2) \times 10^{-4}\ \mathrm{M_{\sun}
yr^{-1}}$. Their result indicates external compression increases the
accretion rate of a protostar, which produces high accretion
luminosity. In this paper, we attribute the origin of higher
luminosity in protostars to star formation activities induced by the
RDI.

Some studies have been made on RDI. \citet{1989ApJ...346..735B} and
\citet{1990ApJ...354..529B} developed an approximate analytical
solution for the evolution of a cloud exposed to ionizing UV
radiation. \citet{1994A&A...289..559L} investigated the RDI with
numerical simulations. However, their calculations did not include
self-gravity of the gas. Recently, SPH simulations including
self-gravity have been done. \citet{2003MNRAS.338..545K}
investigated the effects of initial density perturbations on the
dynamics of ionizing clouds, and \citet{2006MNRAS.369..143M} focused
on cloud morphology. Despite of many studies on RDI, little
attention has been given to the effects of RDI on accretion rates.

In this paper, we explore if the RDI can enhance accretion rates by
orders of magnitude compared with those of protostars that form
without external trigger. RDI is expected to enhance the accretion
rates as a result of strong compression, and it would increase
luminosity of protostars. On the other hand, photo-evaporation of
the parent core due to RDI decreases the mass of protostar,
subsequently decrease the luminosity of the new protostar. We
investigate which effect dominates.

The paper is organized as follows. In Section 2, we describe our
numerical method. In Section 3, we present our results. In Section
4, we discuss luminosities of protostars based on our numerical
results. In Section 5, we summarize our main conclusions.

\section{Numerical Method}\label{sec:method}

We simulate evolution of cores exposed to diffuse UV radiation. We
assumed that a core of neutral gas is immersed within an \ion{H}{ii}
region and is exposed to diffuse radiation from the surrounding
ionized gas. In an actual \ion{H}{ii} region, the core is not only
exposed to diffuse radiation but also to direct radiation from the
ionizing star. The diffuse flux of Lyman continuum photons is $\sim
15 \%$ of the direct flux of Lyman continuum photons
\citep{1998ApJ...502..695C,2004A&A...427..299W}. The spherical core
has an effective cross-section to diffuse radiation $4 \pi r^2$ and
to direct UV radiation $\pi r^2$, respectively. The ratio of total
number of diffuse photons to total number of direct photons is $\sim
0.6$. Our assumption underestimates the total amount of ionizing UV
photons by a factor of $\sim 2.6$. Since the purpose of this paper
is to investigate the effects of RDI on the accretion rate of a
protostar, we neglected direct radiation and considered only diffuse
radiation for simplicity. Effects of direct radiation and morphology
of the core, are beyond the purpose of the present paper, and will
be discussed in our next paper using two dimensional simulations.

The ionizing UV photons enter from outer boundary of the
computational domain. The number flux of these UV photons is
expressed as
\begin{equation}
 F_i = 0.15 \times \frac{N_L}{4 \pi d^2},\label{UV flux}
\end{equation}
where $N_L$ is the total flux of ionizing UV photons from the star,
and $d$ is distance from the ionizing star to the core. We assume
that the ionizing star is an O8V or an O4V star, whose $\log N_L$ is
$48.87 \,{\mathrm s^{-1}}$ and $49.70 \,{\mathrm s^{-1}}$,
respectively \citep{1996ApJ...460..914V}.

We assumed that the sound speed is a function of the ionization
fraction $x$. The hot ionized gas has a temperature of $\sim 10^4\
\mathrm{K}$ in an \ion{H}{ii} region. To imitate condition of the
hot ionized gas ($x=1$), sound speed of the ionized gas $c_i$ was
set to $13 \, \rm{km \, s^{-1}}$. The neutral gas ($x=0$) has two
states: warm gas, if the density is lower than a critical density
$\rho_{crit} = 1.67 \times 10^{-23}\, (10\ m_H)\ \mathrm{g \,
cm^{-3}}$, where $m_H$ is the mass of hydrogen atom, and cold gas,
if the density is higher than the critical density $\rho_{crit}$. We
introduce this artificial condition for the initial pressure balance
between the core and the rarefied warm gas surrounding it. The
rarefied warm gas is ionized in a very short time and will hardly
affect the time evolution of the core. The sound speed of neutral
gas $c_n$ is expressed as
\begin{equation}
 c_n = \left\{ \begin{array}{ll}
    c_{warm} = 10\ \mathrm{km \, s^{-1}} & \mbox{if}\ \rho \leq \rho_{crit} \\
    c_{cold} = 0.2\ \mathrm{km \, s^{-1}} & \mbox{if}\ \rho > \rho_{crit}
           \end{array}\right.,
\end{equation}
where $c_{warm}$ and $c_{cold}$ are the sound speeds of warm gas and cold gas,
respectively. Sound speed of the partially ionized gas ($0<x<1$) is
given as
\begin{equation}
  c = \sqrt{ \left( 1 - x \right) c_n^2 + x c_i^2 }.
\end{equation}

We assume that the initial core is a Bonnor-Ebert sphere, whose mass
$M_0$ is set to 3 $M_{\sun}$ or 30 $M_{\sun}$. A Bonnor-Ebert sphere
is a self-gravitating isothermal sphere in hydrostatic equilibrium
with a surrounding pressurized medium \citep{1956MNRAS.116..351B}.
Gravitational stability of a Bonnor-Ebert sphere is characterized by
a non-dimensional physical truncation radius $\xi_{max}=r_{out}
\sqrt{4 \pi G \rho_c} /c$, where $r_{out}$ is the truncation radius,
$G$ is the gravitational constant, and $\rho_c$ is the central
density. If $\xi_{max}$ is smaller than the critical value
$\xi_{crit}=6.45$, the Bonnor-Ebert sphere is stable. All
Bonnor-Ebert spheres which we use in the calculations are
gravitationally stable ($\xi_{max}=3, 4, 5$).

It is reasonable to assume that the initial density distribution of
the core has spherical symmetry. Density distribution of the core is
assumed to remain unchanged while the core is immersed within an
expanding \ion{H}{ii} region. The sound crossing-time of the
Bonnor-Ebert sphere in model C is $7.02 \times 10^5$ yr, and the
expansion velocity of the \ion{H}{ii} region is approximated well
with $\dot{R_i}=c_i(R_i/R_{st})^{-3/4}$ at $R_i > R_{st}$, where
$R_i$ and $R_{st}$ are the radii of the \ion{H}{ii} region and the
Str\"{o}mgren sphere. If we assume $R_i=10$ pc and an ambient
density of $100\ \mathrm{cm^{-3}}$, the timescale on which the
\ion{H}{ii} region engulfs the core, $2R_0/ \dot{R_i}$, is $4.92
\times 10^4$ yr. This timescale is smaller than the sound-crossing
time, and the adoption of spherical symmetry is acceptable for
simplicity.

We have computed seven models of collapse of cores exposed to UV
radiation from an O star. We summarize the parameters of our
simulations in Table \ref{initial condition}. For the three models
(A, B, and C) we assume that mass of the Bonnor-Ebert sphere is 3
$M_{\sun}$ and the ionizing star is an O8V star. These three models
differ in the non-dimensional physical truncation radii of the
Bonnor-Ebert spheres $\xi_{max}$. Bonnor-Ebert spheres with larger
$\xi_{max}$ have higher central densities and smaller radii. For
other three models (D, E, and F) we assume that the ionizing star is
an O4V star. In model G, we computed the case of a more
massive 30 $M_{\sun}$ core, exposed to UV radiation from an O8V star.

We perform simulations in one spatial dimension using the Godunov
method. Our hydrodynamical code has second-order accuracy in both
space and time. The hydrodynamic equations are
\begin{equation}
  \frac{\partial \rho}{\partial t}
         + \frac{1}{r^2} \frac{\partial r^2 \rho v}{\partial r} = 0,
\label{mass-conservation}
\end{equation}

\begin{equation}
  \frac{\partial \rho v}{\partial t}
         + \frac{1}{r^2} \frac{\partial r^2 \rho v^2}{\partial r} =
         - \frac{\partial P}{\partial r}
         - \frac{GM(r) \rho}{r^2},
\end{equation}

\begin{equation}
  \frac{\partial M}{\partial r} = 4 \pi r^2 \rho,
\label{mass-rho}
\end{equation}
where $r$ is the radius, $\rho$ is the density, $v$ is the radial
velocity, $P$ is the pressure, and $M$ is the mass within  radius
$r$. For near-isothermality, the ratio of specific heats, $\gamma$,
is set to be 1.001. The gas is assumed to be ideal gas.

We solved the radiative transfer equation for the diffuse UV
radiation using a two-stream approximation.
\citet{1998ApJ...502..695C} first treated the diffuse UV radiation
using an iterative two-stream approximation.
\citet{2001A&A...369..263P} simplified the method and showed a
non-iterative version serves as a good approximation for
centrally-peaked density distributions. We adopted the latter for
our Bonnor-Ebert spheres. The equations of evolution for the
ionization fraction and propagation of diffuse UV radiation are
given by
\begin{equation}
  \frac{Dx}{Dt} = - \alpha x^{2}n + 2(1-x) \sigma F(r),
\label{ionization}
\end{equation}

\begin{equation}
  \frac{dF}{dr} = -  2(1-x) n \sigma F(r),
\label{IFpropergation}
\end{equation}
where $n$ is the number density, $F$ is the number flux of ionizing
photons, $\alpha = 2.7 \times 10^{-13} \mathrm{cm^3\, s^{-1}}$ is
the recombination coefficient into the excited state ($n \geq 2$),
and $\sigma = 3.0 \times 10^{-18} \mathrm{cm^2}$ is the ionization
cross-section. These equations were included in the
hydrodynamic solver as follows. First, we solved equations
(\ref{mass-conservation})-(\ref{mass-rho}) by Godunov method, in
which electron density $xn$ is treated as a passively advected
variable. After that, we integrated equations
(\ref{ionization})-(\ref{IFpropergation}) implicitly
\citep{1999MNRAS.310..789W}. Pressure and internal energy were
updated using the new ionization fraction.

We verified our procedure above by a test problem described in
\citet{1994A&A...289..559L}. We simulated the time evolution of
uniform neutral gas illuminated by an ionizing flux increasing
linearly with time in one dimensional Cartesian coordinate. An
ionization front propagates with constant velocity, and neutral gas
accumulated ahead of the ionization front. We adopted same
parameters as \citet{1994A&A...289..559L}. Deviations between
numerical results and the analytic solutions were less than 1 \% and
5 \% for the density and the velocity, respectively. Moreover, we
checked the accuracy of our code in spherical coordinates by
comparing with the self-similar solution of
\citet{1977ApJ...214..488S}.   

The sink cell method \citep[e.g.][]{1982ApJ...258..270B} was used to
avoid the long computational time encountered near the center. As
the gravitational collapse proceeds, the infall velocity increases
at central region and the time step that satisfies the
Courant-Friedrichs-Levy (CFL) condition, which is necessary for
numerical stability, reduces. If the central density exceeds the
reference density $\rho_{sink}\equiv 10^8 \rho_c$, the central 10
grids within 10 AU will be treated as sink cells. The mass that
enters the sink cell is treated as a point mass located at $r=0$.
Boundary condition at the outer boundary of the sink cell was given
by extrapolation from the neighboring cells. Since the infall
velocity near the sink cell is supersonic, this method does affect
the flow of outer gas.

We used non-uniform grids in order to obtain high resolution at the
central region. Sizes of the $i$th grid from center were given as
$\Delta r_{i}=(1+\epsilon) \Delta r_{i-1}$, where $\epsilon$ is
0.02, and size of the most inner grid is 1 AU. The outer boundary is
located at $3 r_{out}$. For example, we allocate 1473 grids in model
C. For the core itself, 1200 grids were allocated. Reflecting and
free boundary conditions were imposed at the inner and outer
boundaries, respectively.

   \begin{table}
      \caption[]{Parameters adopted in the simulations.}\label{initial condition}

         \begin{tabular}{cccccc}
            \hline
        \hline
            Model & $\xi_{max}$ $^a$&  $\rho_c$ $^b$ &  $M_0$ $^c$ &
           spectral type of & d $^d$ \\
       & & [$\mathrm{g \, cm^{-3}}$] &  [$\mathrm{M_{\sun}}$] & ionizing
      star & [pc]  \\
            \hline
            \noalign{\smallskip}
            A & 3 & $1.05 \times 10^{-20}$  & 3 & O8V & $3-10$ \\
            B & 4 & $3.01 \times 10^{-20}$  & 3 & O8V & $3-10$ \\
            C & 5 & $6.09 \times 10^{-20}$  & 3 & O8V & $3-10$ \\

            D & 3 & $1.05 \times 10^{-20}$  & 3 & O4V & $3-10$ \\
            E & 4 & $3.01 \times 10^{-20}$  & 3 & O4V & $3-10$ \\
            F & 5 & $6.09 \times 10^{-20}$  & 3 & O4V & $3-10$ \\

            G & 5 & $6.09 \times 10^{-22}$  & 30 & O8V & $3-10$ \\
            \hline
         \end{tabular}
    $^a$ Non-dimensional truncation radius of initial core (Bonnor-Ebert sphere).\\
    $^b$ Central density of initial core.\\
    $^c$ Total mass of initial core. \\
    $^d$ Distance from ionizing star.
   \end{table}

\section{Results}

\subsection{Time Evolution}\label{subsec:evolution}
\begin{figure*}[t]
    \begin{center}
     \includegraphics[width= 0.47 \textwidth]{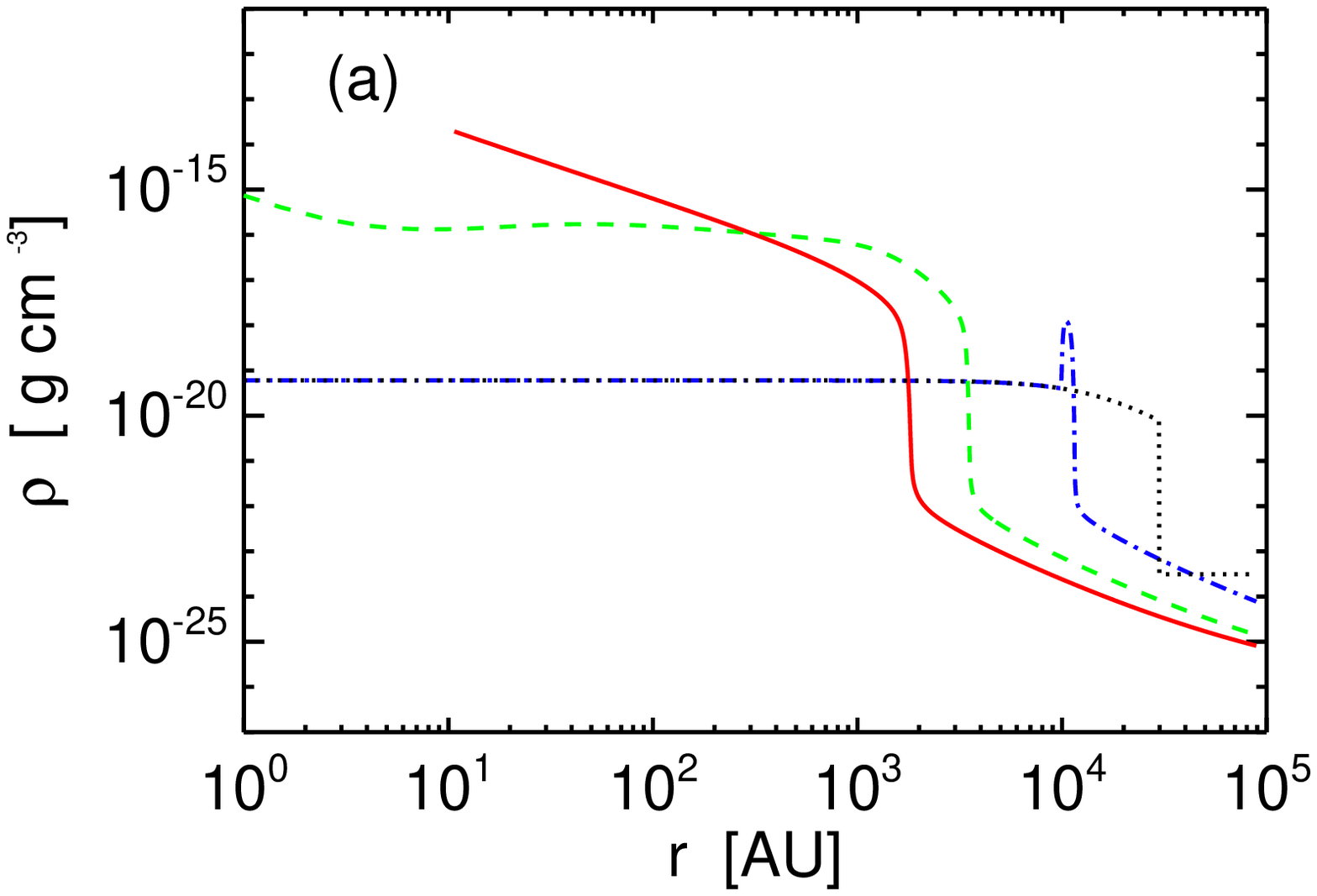}
     \includegraphics[width= 0.47 \textwidth]{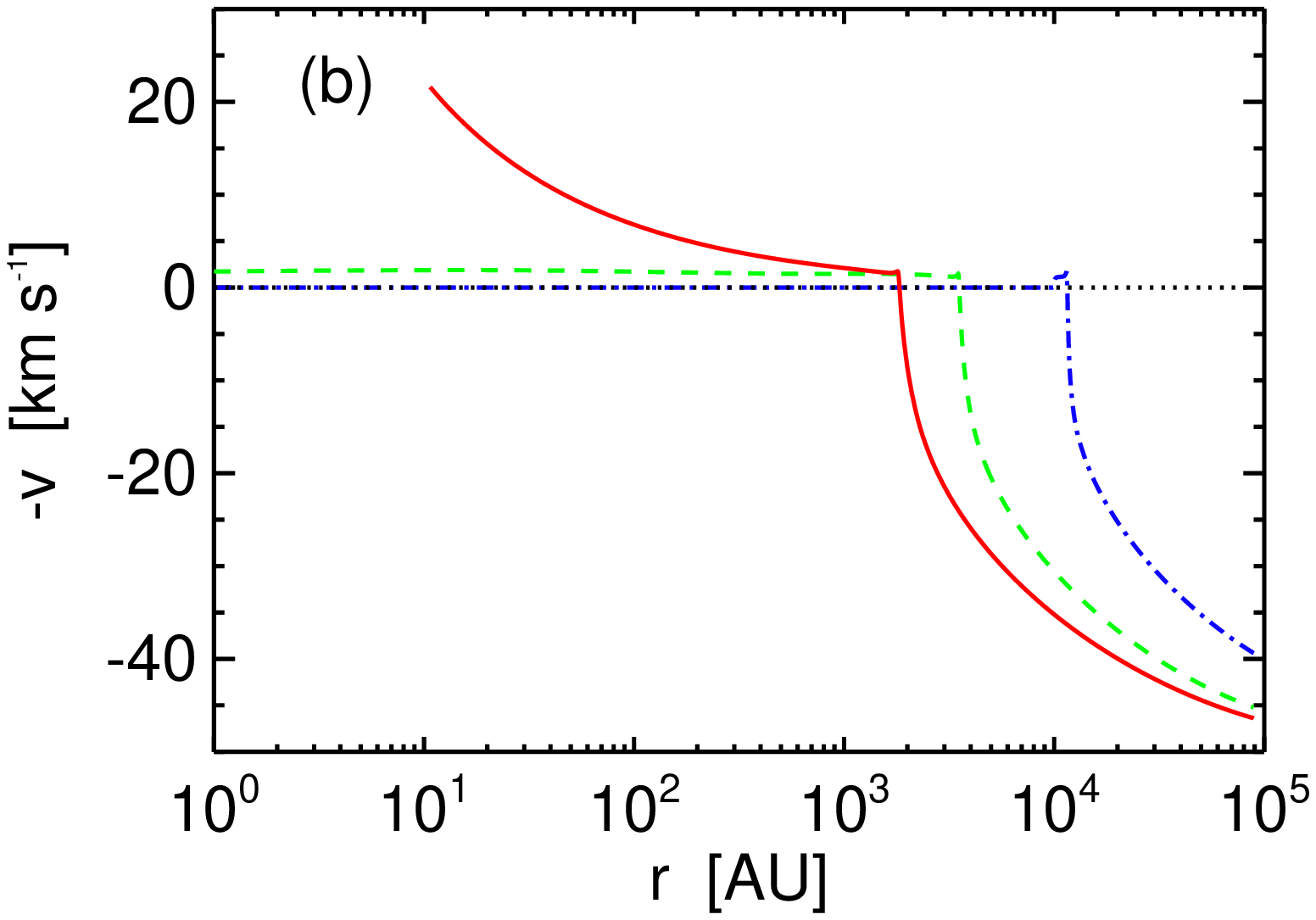}\\
     \includegraphics[width= 0.47 \textwidth]{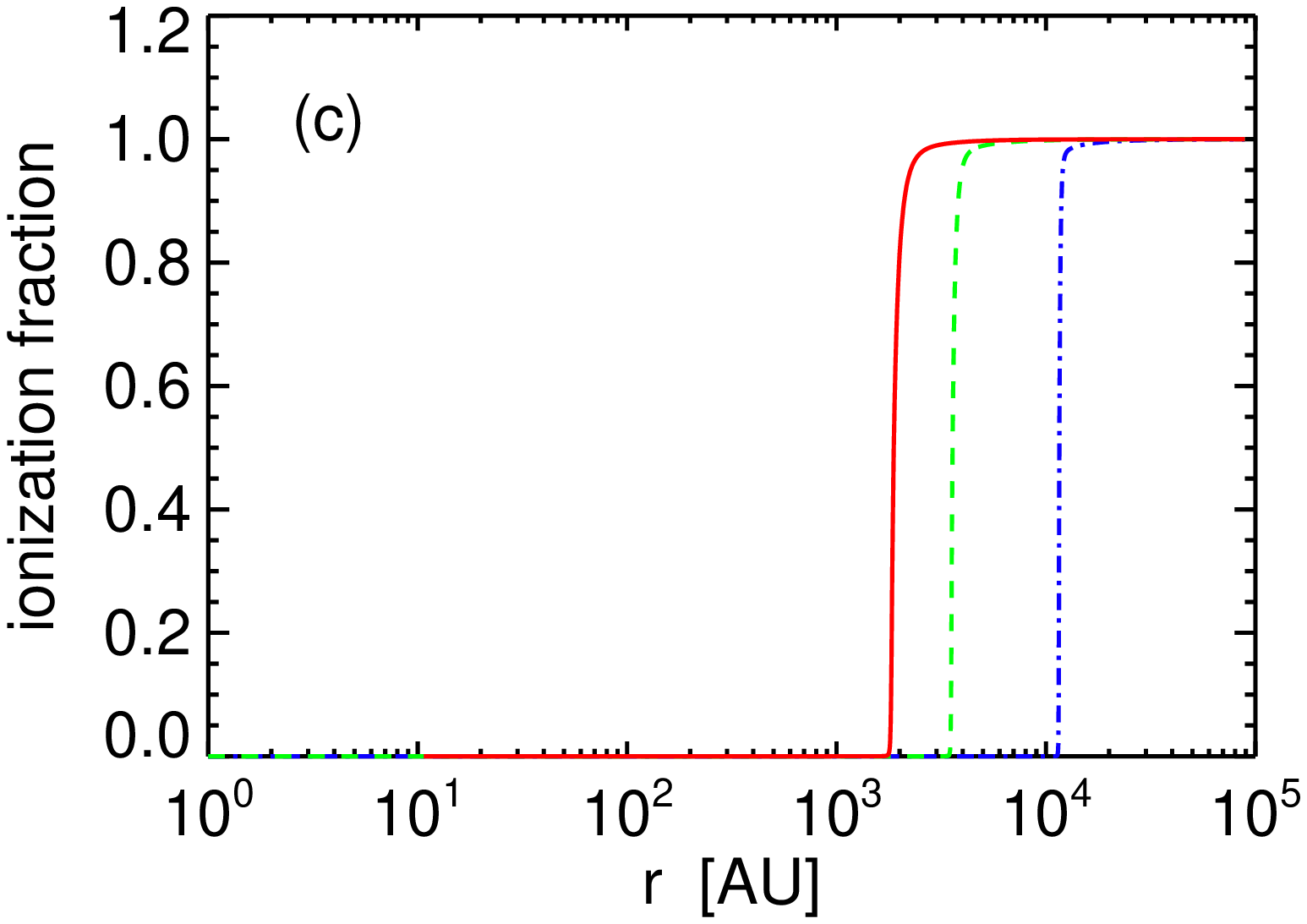}
     \includegraphics[width= 0.47 \textwidth]{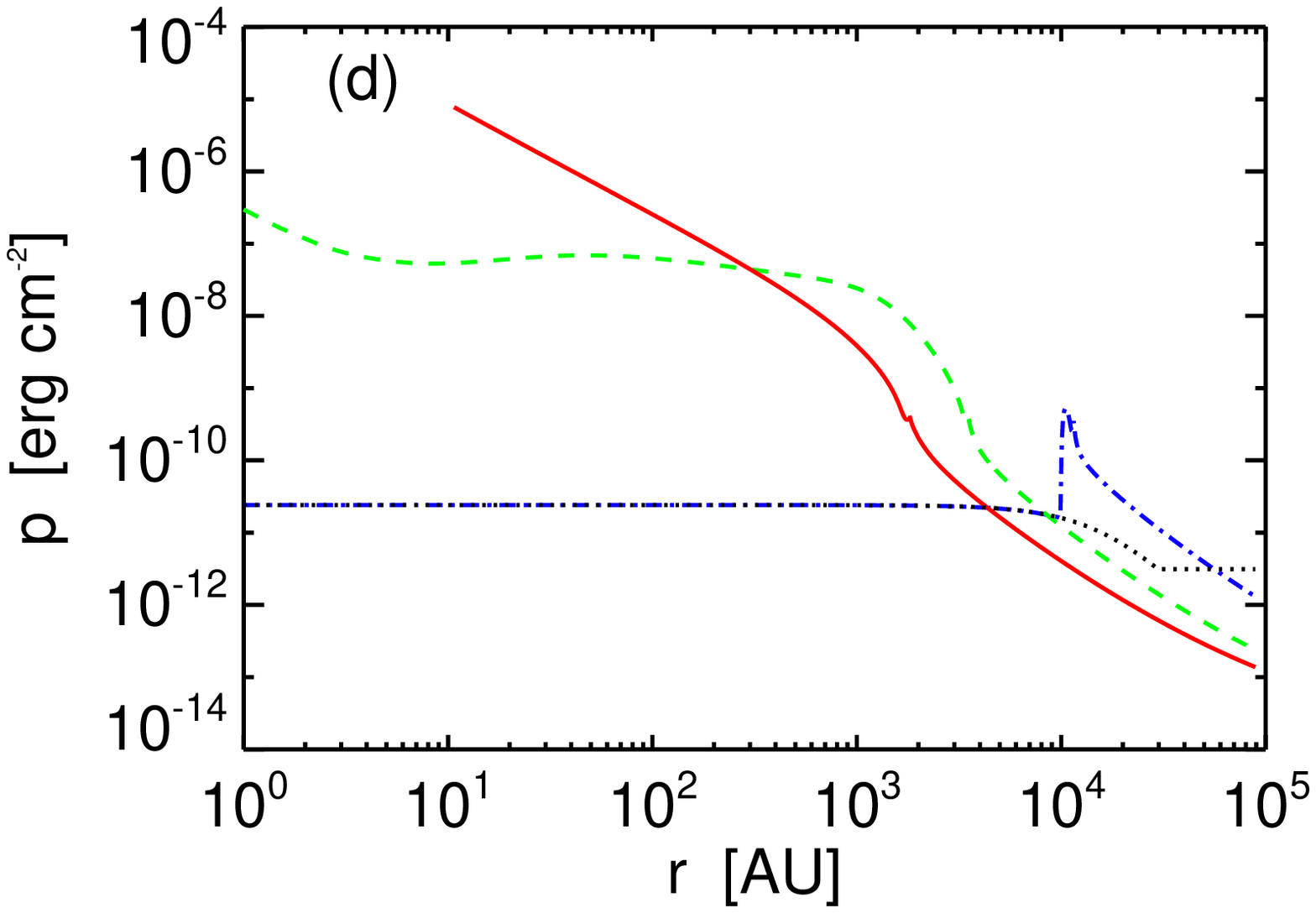}
      \caption{Time evolutions of (a) density profile, (b) velocity profile,
     (c) ionization fraction and (d) pressure profile in the model
     with $\xi_{max}=5$ and $d=9$ pc. The dotted lines denote the
     profiles at $t=0$. The profiles at $t=0.66 \times 10^5 {\mathrm yr}$, $t=1.02 \times
10^5 {\mathrm yr}$, and $t=1.09 \times 10^5 {\mathrm yr}$ are
labeled by dash-dotted, dashed, and solid lines, respectively.}
     \label{fig evolve}
    \end{center}
\end{figure*}

To illustrate common features of the time evolution, we describe
results in model C with $d=9\ \mathrm{pc}$ listed in Table
\ref{initial condition}. Fig. \ref{fig evolve} shows the time
evolution of density, velocity, ionization fraction, and pressure.
Initially, the core was in hydrostatic equilibrium with the rarefied
warm neutral gas. At $t=0$, UV radiation, whose UV photon flux is
$F_i$, turned on and illuminated from the outer boundary. The outer
layer of the core was ionized by UV radiation. Temperature and
pressure at this outer layer increased drastically. The hot ionized
gas then expanded outwards, accelerated to $\sim 40\ \mathrm{km \,
s^{-1}}$, and drove a shockwave into the core. At $t=1.02 \times
10^5\ \mathrm{yr}$, the shockwave reached center of the core. The
dashed line in Fig.\ref{fig evolve}(a) shows that the core is
compressed into a small region of a radius $\sim 3700\ \mathrm{AU}$
at this time. The density becomes $\sim 10^3$ times higher than the
initial density and the core becomes gravitationally unstable.

The core collapses subsequently. We see from Fig.\ref{fig evolve}
(a) and (b) that density and velocity of the compressed core
increase rapidly after $t=1.02 \times 10^5\ \mathrm{yr}$ due to
self-gravity. After the central density exceeds the reference value
$\rho_{sink}$, the sink-cell treatment turns on, at which point a
protostar forms at the center of the core. At this phase, inner
region of the core accretes onto the central protostar, and the
ionized outer layer expands outward.

\begin{figure}[t]
 \centering
 \includegraphics[width=0.5 \textwidth]{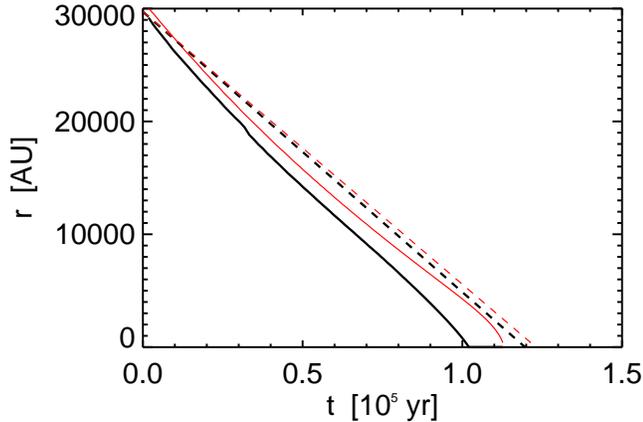}
 \caption{Locations of the ionization front and shock front.
 The thick (black) and thin (red) solid lines label the
 position of shock front and ionization front in model C with
 $d=9\ \mathrm{pc}$, respectively. The thick (black) and
 thin (red) dashed lines label the analytic estimates described in
 section \ref{subsec: analytic}, respectively.}
 \label{fig t_if}
\end{figure}

Velocities of the shock front and the ionization front remain nearly
constant, on the other hand. Fig. \ref{fig t_if} shows locations of
the ionization and the shock fronts as functions of time. Velocity
of the shock front is a constant value of $\sim 1.37\ \mathrm{km\,
s^{-1}}$. Analytic estimation described in section \ref{subsec:
analytic} shows a similar velocity of $1.18\ \mathrm{km\, s^{-1}}$.
For convenience, we define ionization front at the location where
the ionization fraction is 0.9. How ionization front is defined in
fact hardly affects the results, since, as shown in Fig. \ref{fig
evolve}(c), the transition layer from neutral to ionized gas is very
thin. Average velocity of the ionization front is $1.23\
\mathrm{km\, s^{-1}}$, and analytic estimation shows a similar
velocity of $1.14\ \mathrm{km\, s^{-1}}$.

\begin{figure}[t]
 \centering
 \includegraphics[width=0.5 \textwidth]{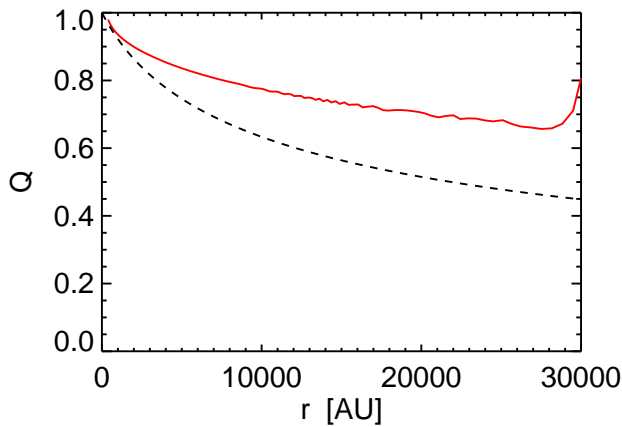}
 \caption{The ratio of UV photon flux at ionization front to
 incident UV photon flux as a function of radius in the model C with
 $d=9\ \mathrm{pc}$. The dashed line
 indicates the analytic estimation which is given by Eq. (\ref{analytic
 Q}).}

 \label{fig r_Q}
\end{figure}

Some fraction of the incident UV photons is absorbed in the layer of
the photo-evaporated gas. In this layer, ionized hydrogen atoms
recombine with electrons at the rate of $\alpha x^{2}n$ per unit
volume per unit time. Some UV photons are used for ionization of
these recombined hydrogen atoms. We define Q, the ratio of UV photon
flux at the ionization front to the incident UV photon flux, as the
photo-evaporation rate of the core. Fig. \ref{fig r_Q} shows Q as a
function of radius. \citet{1978ppim.book.....S} derived an analytic
estimation of $Q$ as
\begin{equation}
 Q = \frac{2}{1 + \sqrt{1 + \frac{4 \alpha F_i r_{IF}}{3 c_i^2}}},
  \label{analytic Q}
\end{equation}
where $r_{IF}$ is the radius of the ionization front. Our numerical
results show a slightly larger value than that of the analytic
estimation by assumptions. \citet{1978ppim.book.....S} assumed that
the photo-evaporated gas expands at the constant sound speed $c_i$.
As shown in Fig.\ref{fig evolve}(b), however, the photo-evaporated gas
accelerated to $\sim 3 c_i \, (\sim 40 \mathrm{k\, ms^{-1}})$ in our
simulations due to pressure gradient. This high velocity reduces the density of the
photo-evaporated gas. As a result, the recombination rate of
hydrogen atoms is smaller than that of the analytic estimate.

\subsection{Accretion Rates}\label{subsec:accretion}
As a reference, we simulate the spontaneous collapse of the core
without UV radiation. Since the thermal pressure balances the
gravitational force in the Bonnor-Ebert sphere, we slightly enhance
the density to $\rho = 1.05\, \rho_{BE}$, where $\rho_{BE}$ is the
density of a marginally stable Bonnor-Ebert sphere ($\xi_{max} =
6.45$), to initiate collapse. This core whose mass is 3 $M_{\sun}$
collapses by self-gravity, and the averaged value of accretion rate
is $7.4 \times 10^{-6}\ \mathrm{M_{\sun}\ yr^{-1}}$.

\begin{figure}[t]
   \centering
   \includegraphics[width=0.5 \textwidth]{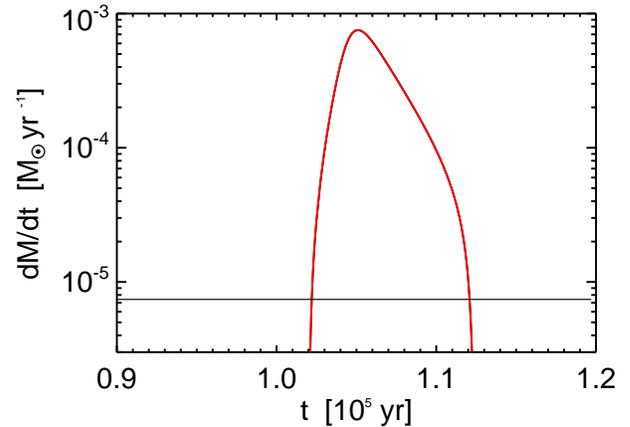}
   \caption{Accretion rate as a function of time in the model C with
    $d=9\ \mathrm{pc}$. The horizontal line
    indicates the averaged value of accretion rate in model without UV
    radiation.}
   \label{fig time-accre}
    \end{figure}

\begin{figure}[t]
   \centering
   \includegraphics[width=0.5 \textwidth]{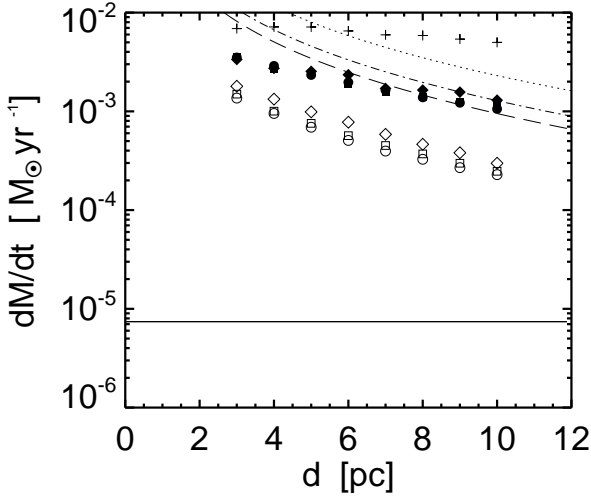}
   \caption{The averaged accretion rates as
    functions of distance from the ionizing star. Open diamonds, open squares,
    open circles label results from the model A,
    B, and C, respectively. Filled diamonds, filled squares, filled circles
    label models D, E, and F, respectively.
    The crosses are for the model G. The horizontal line indicates the averaged
    accretion rate in the reference model. Dotted line, dash-dotted line and
    dashed line indicate analytic estimation for models A, B, and C, respectively.}
   \label{fig dist-accre}
    \end{figure}

The RDI indeed increases accretion rates. Fig. \ref{fig time-accre}
shows the accretion rate as a function of time in model C with $d=9\
\mathrm{pc}$. The accretion rates are measured at the radius of 10
AU. The horizontal line indicates the average accretion obtained
from the reference model. At $t=1.02 \times 10^5\ \mathrm{yr}$, the
central density reaches the threshold density $\rho_{sink}$, and the
sink cell method turns on. The accretion rate increases rapidly
afterwards, and lasts for $\sim 10^4 \mathrm{yr}$.
Fig. \ref{fig dist-accre} shows the averaged values of accretion
rates as a function of distance from the ionizing star. The
accretion rates increase with decreasing distance from the ionizing
star, and with the increasing UV photon flux. If the cores are
located within 10 $\mathrm{pc}$ from an ionizing star, the accretion
rates can go 1-2 orders higher than models without UV radiation.

The accretion rates not only depend on the UV photon fluxes but also
on $\xi_{max}$ and $M_0$. Cores with larger values of $\xi_{max}$ have smaller
radii and higher densities. Shock velocities and densities in
post-shock regions are larger in the models with small $\xi_{max}$.
Gas of higher density has a shorter free-fall time, and reaches an
accretion rate higher in the models with smaller $\xi_{max}$. We will
compare this with analytic estimates for better understanding. In the
model G, not only small
free-fall time but also large mass of compressed core is responsible for high
accretion rate.

\subsection{Final Masses}\label{subsec:final mass}

Photo-evaporation can reduce the actual mass that accretes onto the
protostar. In the outer layer, hot ionized gas expands outward at a
velocity up to $\sim 40\ \mathrm{km \, s^{-1}}$. Photo-evaporation
of gas takes place at $\sqrt{2 G M / r}$, except for the case of
very small cores. The escape velocity of the Bonnor-Ebert sphere in
model C is $\sim 0.42\ \mathrm{km \, s^{-1}}$, for example, is much
smaller than that of the photo-evaporated flow. Photo-evaporation
prevents large amount of gas from accreting onto the protostar, and
final mass of the protostar can be small due to mass loss by
photo-evaporation.

\begin{figure}[t]
   \centering
   \includegraphics[width=0.5 \textwidth]{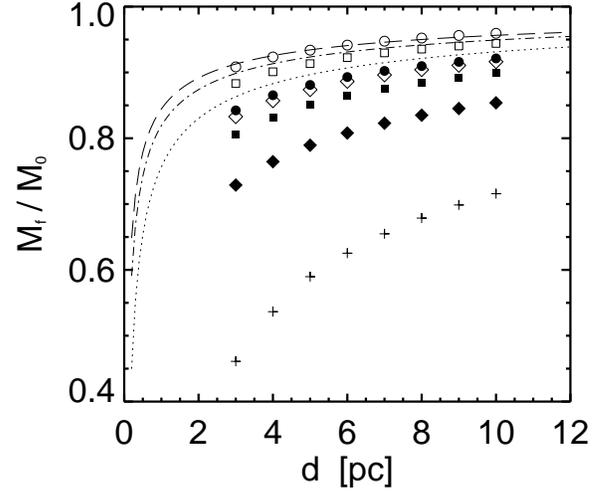}
   \caption{Ratio of final mass of protostar to initial core mass as a
    function of distance from ionizing star. The open  diamonds, the
    open squares, the open circles indicate results in the models A,
    B, and C, respectively. The filled  diamonds, the filled squares, the
    filled circles indicate results in the models D, E, and F,
    respectively. The crosses indicate results in the model G. The
    dotted line, the dash-dotted line and the dashed line indicate
    analytic estimation for model A, B,and C, respectively.}
   \label{fig dist-mass}
    \end{figure}

Final masses of protostars decrease with decreasing distance from
the ionizing star. Fig. \ref{fig dist-mass} shows the final masses
of protostars as functions of distance from ionizing star. The mass
of a protostar is defined as mass which has entered sink cells, and
calculations continue until accretion rates become sufficiently
small ($10^{-9}\ \mathrm{M_{\sun}\,yr^{-1}}$). Mass loss due to
photo-evaporation at the surface of the core is proportional to the
incident UV photon flux and final masses of protostars are smaller
near the ionizing star.  For an O8V or an O4V star, mass loss due to
photo-evaporation varies from several percent to several tenth of
percent of the initial core mass. The more massive the ionizing star
gets, the smaller the final mass.

The final masses not only depend on the UV photon fluxes but also
on the sizes of cores. Fig. \ref{fig dist-mass} shows that mass loss
due to photo-evaporation
is significant in the model G, in which initial radius of the core is 10
times larger than that in the model C. Larger core has
larger photo-evaporation rate and longer time to be exposed to UV radiation
because of larger surface area and longer crossing time of shock
wave. This is the reason why the core lose large amount of
mass in the model G. For the same reason, final masses are small in the
models with small $\xi_{max}$.

\section{Discussion}
\subsection{Analytic Estimates}\label{subsec: analytic}
In this section, we derive analytic estimates of final mass and
accretion rates of protostars under the influence of RDI. For
simplicity, we assume that a core has a uniform density $n_0$ and a
radius $R_0$ initially. The core is exposed to UV radiation whose
number flux of UV photons is $F_i$. Spherical symmetry is assumed.

The final mass of protostar $M_f$ is expressed as
\begin{equation}
 M_f = M_0 - \int_{0}^{t_s}\! 4 \pi r_{IF}(t)^2 \dot{m}\, dt,
\label{final mass}
\end{equation}
where $r_{IF}(t)$ is the radius of the ionization front at time $t$,
$\dot{m}$ is the mass loss flux at radius $r_{IF}(t)$, and $t_s$ is
the crossing time of the shockwave. The second term on the
right-hand side of Eq. (\ref{final mass}) expresses mass loss due to
photo-evaporation. We only consider mass loss during the compression
phase when the core is compressed into a small region by the shock.
After the compression, the core collapses within the free-fall time
of the compressed core that is much shorter than $t_s$. We need
$\dot{m}$, $t_s$, and $r_{IF}(t)$ to estimate the final mass of
protostars using Eq. (\ref{final mass}).

The mass loss flux $\dot{m}$ due to photo-evaporation is expressed
as
\begin{equation}
 \dot{m} = m_H Q F_i.
  \label{m dot}
\end{equation}
where $Q$ is the ratio of UV photon flux which arrives at surface of
core to incident UV photon flux $F_i$.  The UV photons ionize the
outer layer of the core, but some fraction gets absorbed in the
photo-evaporated flow. As shown in Fig. \ref{fig r_Q}, $Q$
approaches to unity as the ionization front enters the inner region
of the core. However, we assume that $Q$ remains constant through
the compression phase and calculate the value of $Q$ at the radius
$R_0$ using Eq. (\ref{analytic Q}).

The shock velocity $u_s$ is derived following
\citet{1989ApJ...346..735B} to estimate the shock crossing time
$t_s$. Assuming ram pressure of the flow equals thermal pressure of
the post-shock region, we obtain
\begin{equation}
  n_0 u_s^2 = c_n^2 n_1,
\label{P balance at SF}
\end{equation}
where $n_0$ and $n_1$ are number densities of the undisturbed region
and the post-shock region, respectively. The equation of mass
continuity at the ionization front gives
\begin{equation}
 Q F_i = n_1 u_{IF}',
\label{balance at IF}
\end{equation}
where $u_{IF}'$ is the velocity of ionization front in the rest
frame of the post-shock region. The ionization front is assumed to
be D-critical \citep[cf.][]{1978ppim.book.....S}, and velocity of
the ionization front is given as
\begin{equation}
 u_{IF}' \simeq \frac{c_n^2}{2 c_i}.
\label{Dcritical}
\end{equation}
Substituting Eq. (\ref{Dcritical}) into Eq. (\ref{balance at IF}) gives
\begin{equation}
 2 Q c_i F_i = c_n^2 n_1.
\label{balance at IF 2}
\end{equation}
Combining Eq. (\ref{P balance at SF}) and Eq. (\ref{balance at IF 2}), we obtain
\begin{equation}
 u_s = \sqrt{\frac{2 c_i Q F_i}{n_0}}.
  \label{shock velocity}
\end{equation}
Using Eq. (\ref{shock velocity}), the shock crossing time is
expressed as
\begin{equation}
 t_s = \frac{R_0}{\sqrt{\frac{2 c_i Q F_i}{n_0}}}.
  \label{shock crossing time}
\end{equation}

We derive the velocity of ionization front $u_{IF}$ in the rest
frame of the undisturbed region to obtain radius of the ionization
front $r_{IF}(t)$. Taking into account $n_1 = \left( \frac{u_s}{c_n}
\right)^2 n_0$ and $n_0 u_s = n_1 (u_s -v_1)$, we obtain
\begin{equation}
 v_1 = u_s - \frac{c_n^2}{u_s},
\label{velocity ps region}
\end{equation}
where $v_1$ is velocity of the post-shock region in the rest frame
of the undisturbed region. Velocity of the ionization front is
\begin{equation}
 u_{IF} = u_s - \frac{c_n^2}{u_s} + \frac{c_n^2}{2 c_i}.
\end{equation}
Using $u_{IF}$, radius of the ionization front is expressed as
\begin{equation}
   r_{IF} = R_0 - u_{IF} t.
\label{r_if}
\end{equation}

The final mass of protostar using Eq. (\ref{final mass}) gives
\begin{equation}
 M_f = M_0 - 4 \pi m_H Q F_i\, (R_0^2 t_s - R_0 u_{IF} t_s^2 +
      \frac{u_{IF}^2}{3} t_s^3).
      \label{final mass 2}
\end{equation}
This equation gives us the analytic estimate of final mass of a
protostar influenced by RDI.

The accretion rate, over a free-fall time, is roughly estimated by
\begin{equation}
\dot{M} =\frac{M_{f}}{t_{ff}},
 \label{estimation of accretion}
\end{equation}
where $t_{ff} = \sqrt{\frac{3 \pi}{32 \rho_f G}}$ is the free-fall
time of the compressed core. The core is compressed into a radius of
$r_{IF}(t_s)$ when the shock reaches its center. The density of the
compressed core is then given as
\begin{equation}
 \rho_f = \frac{M_f}{\frac{4 \pi}{3} r_{IF}(t_s)^3},
\end{equation}
and Eq. (\ref{estimation of accretion}) gives us the analytic
estimate of the accretion rate.

\subsection{Luminosity}\label{subsec: luminosity}
Luminosity of a protostar is mainly produced by accretion, because
nuclear burning does not influence much of the luminosity during the
protostellar collapse phase. The accretion luminosity $L$ is
expressed as
\begin{equation}
 L =\frac{G M_{*} \dot{M}}{R_*},
   \label{luminosity}
\end{equation}
where $M_*$ is the mass, $\dot{M}$ is the accretion rate, and $R_*$
is the radius of the protostar. Enhancement of accretion rates due
to RDI has a large effect on the luminosity. Although the accretion
rate depends not only on the distance from the ionizing star but
also on the $\xi_{max}$ of the Bonnor-Ebert sphere, it is relatively
not as sensitive to $\xi_{max}$ compared with the distance. Our
results show that for cores located within 10 pc of an ionizing
star, the accretion rates becomes 1-2 orders of magnitude higher
than those without triggering from the UV radiation.

Enhanced luminosity around bright-rimmed clouds is seen in
observations. \citet{1989ApJ...342L..87S} observed three
bright-rimmed globules associated with cold IRAS sources in
$^{13}$CO(J=1-0) emission.  They compared the physical parameters with four
isolated dark globules and found that the ratios of the luminosity
to globule mass, $3-13\ L_{\sun}/M_{\sun}$, are 1-2 orders higher
than the values of isolated dark globules, $0.03-0.3\
L_{\sun}/M_{\sun}$, as shown by our calculations. These objects are
good candidates influenced by the RDI.

Protostars near \ion{H}{ii} regions may spend most of their lifetime
in the state of high luminosity. If the duration of high accretion
rate is much shorter than the lifetime of the protostar, it is
relatively difficult to catch the protostar in its high-luminosity
state by observations. In the simulations, the protostellar phase
begins from the time when the sink-cell method is turned on and
lasts until 99 \% of the final mass has accreted within the sink
cells. The lifetime of the protostar in model C is  $9.1 \times
10^3\ \mathrm{yr}$ with $d=9\ \mathrm{pc}$. As shown in Fig.
\ref{fig time-accre}, the accretion rate is enhanced for a duration
of $8.4 \times 10^3\ \mathrm{yr}$, comparable to the lifetime of the
protostar. It is likely that protostars near \ion{H}{ii} regions be
observed in the state of higher luminosity.

\subsection{Comparison with observations}
   \begin{table*}
    \caption[]{Star formation regions of potential triggered origins}
    \label{other regions}
    \begin{center}
    \begin{tabular}{cccccc}
     \hline
     \hline

     Cloud      &  Spectral type of & Projected distance from &
     Estimated shock &  &
     Evidence of triggered \\
     & main ionizing star & ionizing star &
     velocity & & star formation \\
     \hline
     molecular pillars in M16$^1$  &  O5V  & 2 pc  & 1.3 $\mathrm{k\, ms^{-1}}$ &
     & age gradient of YSOs \\

     IC1396N$^2$  &  O6.5f & 11 pc & 0.6 $\mathrm{k\, ms^{-1}}$& &
     age gradient of YSOs \\

     SFO11, SFO11NE, SFO11E$^3$  &   O6V  & 11 pc & 1.4 $\mathrm{k\, ms^{-1}}$&
     & on-going star formation\\
     & & &  & & near ionization front  \\

     Orion bright-rimmed clouds$^4$  & O8III$^a$ or O6:$^b$  & - &  -   &
     & on-going star formation \\
     & & &  & & near ionization front  \\

     \hline
    \end{tabular}
     \end{center}
    $^a$  Ionizing star of B30 and B35. \\
    $^b$  Ionizing star of IC2118, LDN1616, and LDN1634. \\
    References: (1) \citet{2002ApJ...568L.127F},
    \citet{1999A&A...342..233W}, (2) \citet{2007ApJ...654..316G}, (3)
    \citet{2004A&A...414.1017T}, (4) \citet{2005ApJ...624..808L}
   \end{table*}

Evidences of triggered star formation have been suggested on edges
of many \ion{H}{ii} regions. Some of the observations in fact favor
the scenario of sequential star formation around a massive star.
\citet{2005ApJ...624..808L} selected candidates of pre-main sequence
stars based on 2MASS colors in the Orion region, and revealed
spatial distribution of classical T Tauri stars using their
spectroscopic observations. The pre-main sequence stars form between
the ionizing star and the bright-rimmed clouds, but not deeply
inside the cloud. The youngest stars are found near the cloud
surfaces, forming an apparent age gradient. As the \ion{H}{ii}
region expands, it might be able to trigger star formation
sequentially from near to far. Several of such examples are
summarized in Table 2.

RDI may contribute to the triggering mechanism around a massive
star. \citet{2005ApJ...624..808L} found that only bright-rimmed
clouds with strong IRAS 100 $\mu m$ and H$\alpha$ emission show
signs of ongoing star formation. This suggests that star formation
may have been triggered by strong compression from the ionization
shock, because these emissions are tracers of dense gas and
ionization.  Small-scale age gradients of YSOs near the ionization
fronts are seen in some molecular clouds.
\citet{2002ApJ...568L.127F} observed heads of molecular pillars
$\Pi_1$ and $\Pi_2$ in the Eagle Nebula (M16).  They found that
protostars and starless cores are ordered in age sequences. The more
evolved objects are closer to the cloud heads (and the ionizing
star) and the younger objects are farther from the exciting star.
Similar age gradients are also seen in bright-rimmed clouds
\citep{1995ApJ...455L..39S} and cometary globule IC1396N
\citep{2007ApJ...654..316G}.  These age gradients near ionization front
seem to suggest that star formation may be sequentially triggered by
ionization shocks.

Stars formed facing the \ion{H}{ii} regions appear to be more
luminous. \citet{1999PASJ...51..791Y} carried out $ ^{13}
\mathrm{CO(J=1-0)}$ observations toward 23 southern \ion{H}{ii}
regions and identified 95 molecular clouds associated with
\ion{H}{ii} regions. They found that IRAS point sources tend to be
more luminous on the side facing an \ion{H}{ii} region. Their result
is consistent with enhanced luminosity as a result of the RDI.

Shock speeds derived from observations are similar.
\citet{1999A&A...342..233W} and \citet{2004A&A...414.1017T} derived
shock velocities from pressure in pre- and post-shock regions.
\citet{2007ApJ...654..316G} derived shock velocity from the age
gradients of YSOs. Their estimated values of a few $\mathrm{km\,
s^{-1}}$ are consistent with our results. For example,
\citet{2004A&A...414.1017T} estimated that shock velocity in
bright-rimmed cloud SFO11 is $\sim 1.4\ \mathrm{km\,s^{-1}}$. SFO11
is illuminated by an O6V star whose projected distance is 11 pc.
These conditions are close to those in model C with $d=9$, whose
ionizing star is an O8V star. As described in section
\ref{subsec:evolution}, the averaged shock velocity is $\sim 1.37\
\mathrm{km\, s^{-1}}$ in this model.

\citet{2006Lee} found that classical T Tauri stars exist near the
surfaces of bright-rimmed clouds and intermediate mass stars exist
relatively inside clouds. One possible difference of stellar mass
may be density or mass in the parent cores. Photo-evaporation may
reduce the mass that would otherwise accrete onto the stars. In our
simulations a few percent to several tenth percent of initial mass
may be lost by photo-evaporation. Mass of young stars may be smaller
near the surface of bright-rimmed clouds compared with inner
regions.

\section{Conclusions}
We simulate the evolution of cores exposed to strong UV radiation
around the \ion{H}{ii} regions.  We investigate possible effects of
the radiation-driven implosion on the accretion luminosity of
forming protostars. We also developed analytic estimates of
accretion rates and final masses of protostars, which are compared
with the numerical results.

The main findings are:
   \begin{enumerate}
    \item Radiation-driven implosion can compress the cores
      and decrease the free-fall time.
    \item The accretion rates are positively influenced by on the incident UV
      photon flux.
    \item Photo-evaporation of the parent cores may decrease the final masses of
      protostars. However, mass loss due to photo-evaporation does not
      affect significantly the accretion luminosity.
    \item Final mass of a protostar depends mainly on the size of
      the parent core and incident UV photon flux. Final mass of a protostar
      decreases with the increase of UV photon flux. Dense compact
      cores, on the other hand, are hardly affected by mass loss due to RDI.
    \item The RDI increases luminosity of protostars
    by orders of magnitude than spontaneous star formations without RDI.
   \end{enumerate}


\begin{acknowledgements}
The authors would like to thank Wen-Ping Chen, Huei-Ru Chen,
Jennifer Karr, and Ronald Taam for their helpful discussions. This
work is supported by the Theoretical Institute for Advanced Research
in Astrophysics (TIARA) operated under Academia Sinica, and the
National Science Council Excellence Projects program in Taiwan
administered through grant number NSC95-2752-M-001-008-PAE, and
NSC95-2752-M-001-001. Numerical computations were in part carried
out on VPP5000 at the Astronomical Data Analysis Center, ADAC, of
the National Astronomical Observatory of Japan.
\end{acknowledgements}

\bibliographystyle{aa} 
\bibliography{references} 

\end{document}